\documentclass[aps,prl,amsmath,amssymb,reprint,preprint,preprintnumbers,superscriptaddress,showpacs,groupedaddress,lengthcheck]{revtex4-1}
\newcommand{\bscco}{$\textrm{Bi}_2\textrm{Sr}_2\textrm{CaCu}_2\textrm{O}_{8+\delta}$}

\usepackage{graphicx}% Include figure files
\usepackage{dcolumn}% Align table columns on decimal point
\usepackage{bm}% bold math

\begin{document}

%\preprint{}

\title{Scattering from incipient stripe order in the high-temperature superconductor \bscco}


\author{Eduardo H. da Silva Neto}
%\thanks{}
\affiliation{Joseph Henry Laboratories and Department of Physics, Princeton University, Princeton, NJ 08544, USA}

\author{Colin V. Parker}
\affiliation{Joseph Henry Laboratories and Department of Physics, Princeton University, Princeton, NJ 08544, USA}

\author{Pegor Aynajian}
\affiliation{Joseph Henry Laboratories and Department of Physics, Princeton University, Princeton, NJ 08544, USA}

\author{Aakash Pushp}
\altaffiliation{Current Address: IBM Almaden Research Center, 650 Harry Road, San Jose, CA 95120, USA}
\affiliation{Joseph Henry Laboratories and Department of Physics, Princeton University, Princeton, NJ 08544, USA}

\author{Jinsheng Wen}
\author{Zhijun Xu}
\affiliation{Condensed Matter Physics and Materials Science, Brookhaven National Laboratory (BNL), Upton, NY 11973, USA}

\author{Genda Gu}
\affiliation{Condensed Matter Physics and Materials Science, Brookhaven National Laboratory (BNL), Upton, NY 11973, USA}

\author{Ali Yazdani}
\affiliation{Joseph Henry Laboratories and Department of Physics, Princeton University, Princeton, NJ 08544, USA}
\email{yazdani@princeton.edu}


\date{\today}

\begin{abstract}
Recently we have used spectroscopic mapping with the scanning tunneling microscope to probe modulations of the electronic density of states in single crystals of the high temperature superconductor \bscco\ (Bi-2212) as a function of temperature [C. V. Parker \textit{et al}., Nature (London) \textbf{468}, 677 (2010)]. These measurements showed Cu-O bond-oriented modulations that form below the pseudogap temperature with a temperature-dependent energy dispersion displaying different behaviors in the superconducting and pseudogap states. Here we demonstrate that quasiparticle scattering off impurities does not capture the experimentally observed energy- and temperature-dependence of these modulations. Instead, a model of scattering of quasiparticles from short-range stripe order, with periodicity near four lattice constants ($4a$), reproduces the experimentally observed energy dispersion of the bond-oriented modulations and its temperature dependence across the superconducting critical temperature, $T_c$. The present study confirms the existence of short-range stripe order in Bi-2212.
\end{abstract}

\pacs{74.72.Gh, 74.55.+v, 74.20.Pq}
%\keywords{}

\maketitle

%\section{}
%\subsection{}
%\subsubsection{}

Understanding the interplay between superconductivity, the pseudogap \cite{timusk_pseudogap_1999}, and the possibility of charge and spin ordering phenomena in the proximity to the Mott insulating ground state continues to be one of the most challenging problems in condensed matter physics. Periodic patterns of spin and charge, referred to as stripe order \cite{zaanen_charged_1989, tranquada_evidence_1995, kivelson_electronic_1998, white_density_1998, vojta_lattice_2009}, have been detected in static form in the La-based cuprates \cite{tranquada_evidence_1995}. Yet their relevance to the wider class of high-$T_c$ compounds and significance to the mechanism of superconductivity and pseudogap remains unknown. Addressing this question is challenging since in the presence of disorder, or in the absence of favorable structural distortion \cite{cava_crystal_1987, buechner_critical_1994}, such an order is likely short range or fluctuating and hence hard to detect \cite{kivelson_how_2003, robertson_distinguishing_2006}. While the observation of static stripe patterns with scattering techniques is well established, detecting fluctuating order and distinguishing it from other electronic spatial modulations is still being developed \cite{kivelson_how_2003}. Establishing the presence of fluctuating or short-range stripe order in the cuprates and understanding its correlation with other phenomena, such as superconductivity or the pseudogap, is of great importance.

In this paper we analyze spectroscopic measurements with the STM that allow us to probe the spatial variation of the local density of states (LDOS) in single crystals of Bi-2212. Recently, we have shown that spatial features of the LDOS associated with incipient stripe order \cite{parker_fluctuating_2010} can be distinguished from those due to impurity-induced quasiparticle interference \cite{hoffman_imaging_2002, kohsaka_how_2008, hanaguri_quasiparticle_2007, wise_imaging_2009} in this compound. These previous measurements demonstrate that signatures of incipient stripe order, which appear as bond-oriented modulations in the LDOS \cite{howald_coexistence_2003, vershinin_local_2004, kohsaka_intrinsic_2007, wise_charge-density-wave_2008}, first appear below the pseudogap temperature $T^*$ across a wide range of doping.  We have also shown these modulations in Bi-2212 to have the strongest intensity when the samples’ hole doping is close to $1/8$, the concentration at which static stripes have been found in La-based cuprates \cite{tranquada_evidence_1995} and as predicted by most stripe models \cite{kivelson_electronic_1998, white_density_1998, vojta_lattice_2009, kivelson_how_2003}. Yet, the wavelength associated with these modulations shows a significant energy-dependence, a behavior not expected for static long-range order. Furthermore, the energy-momentum structure of these modulations in near optimally doped samples exhibits dramatically different behaviors across the superconducting transition temperature $T_c$. Here we show that a model of scattering of quasiparticles from short range stripe order can capture the energy-dependence of the LDOS modulations both above and below $T_c$.  This model also provides an insight into the particle-hole symmetry of the electronic states in the superconducting and pseudogap phases. These findings demonstrate the strong interplay between incipient stripe order and the electronic properties of both superconducting and pseudogap states of the cuprates.  Establishing this connection in compounds other than the La-based compounds suggests that incipient stripe order plays an important role in all families of cuprates.

\begin{figure*}
\includegraphics[width=145mm]{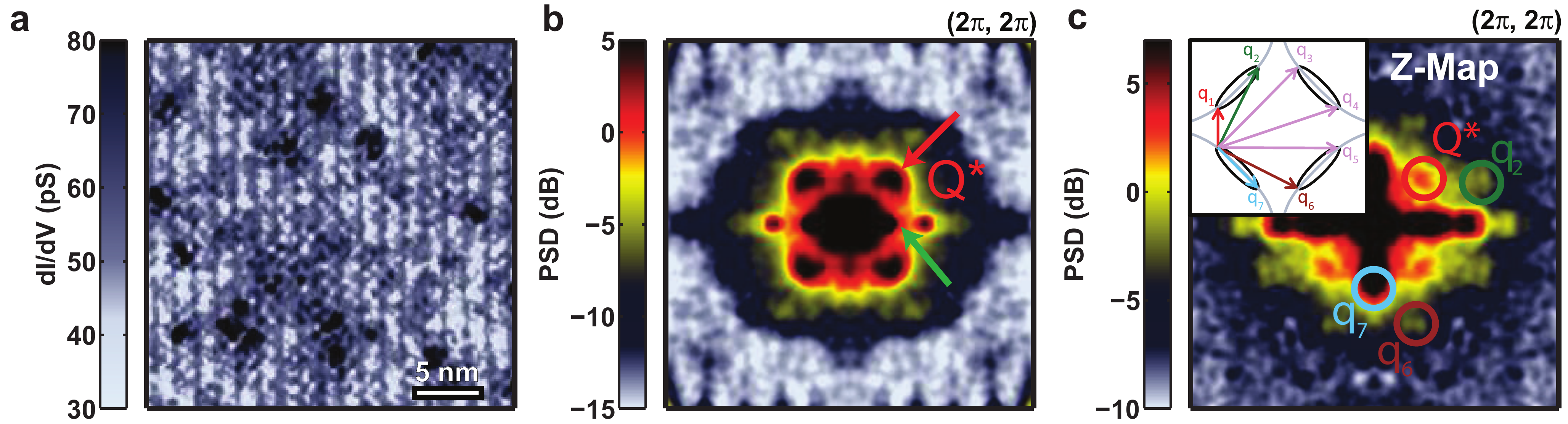}
\caption{\label{fig:1} Optimally doped Bi-2212 sample ($T_c = 91$ K). (a) Real space mapping of the low-energy conductance, $G(\bm{r}, V = 10$ mV$)$, taken at 35 K, and measured at a setup bias of $-150$ mV and setup current of $40$ pA. (b) DFT of the data in (a), showing strong peaks corresponding to the Cu-O bond oriented modulations marked $Q^*$ (red arrow) and b-axis supermodulation (green arrow). (c) DFT of $Z(\bm{r}, V = 10$ mV$)$ on the same area and temperature. Inset shows schematic of the octet model predicting seven wavevectors ($q_1$ through $q_7$), four of which can be seen in the DFT in (c). Corner of the DFT maps (a,b) correspond to $(2\pi, 2\pi)$ in units of $1/a_0$, where $a_0=\sqrt{2} a$ and $a$ is the nearest neighbor Cu-Cu distance. PSD stands for power spectral density.}
\end{figure*}

\begin{figure}
\includegraphics[width=75mm]{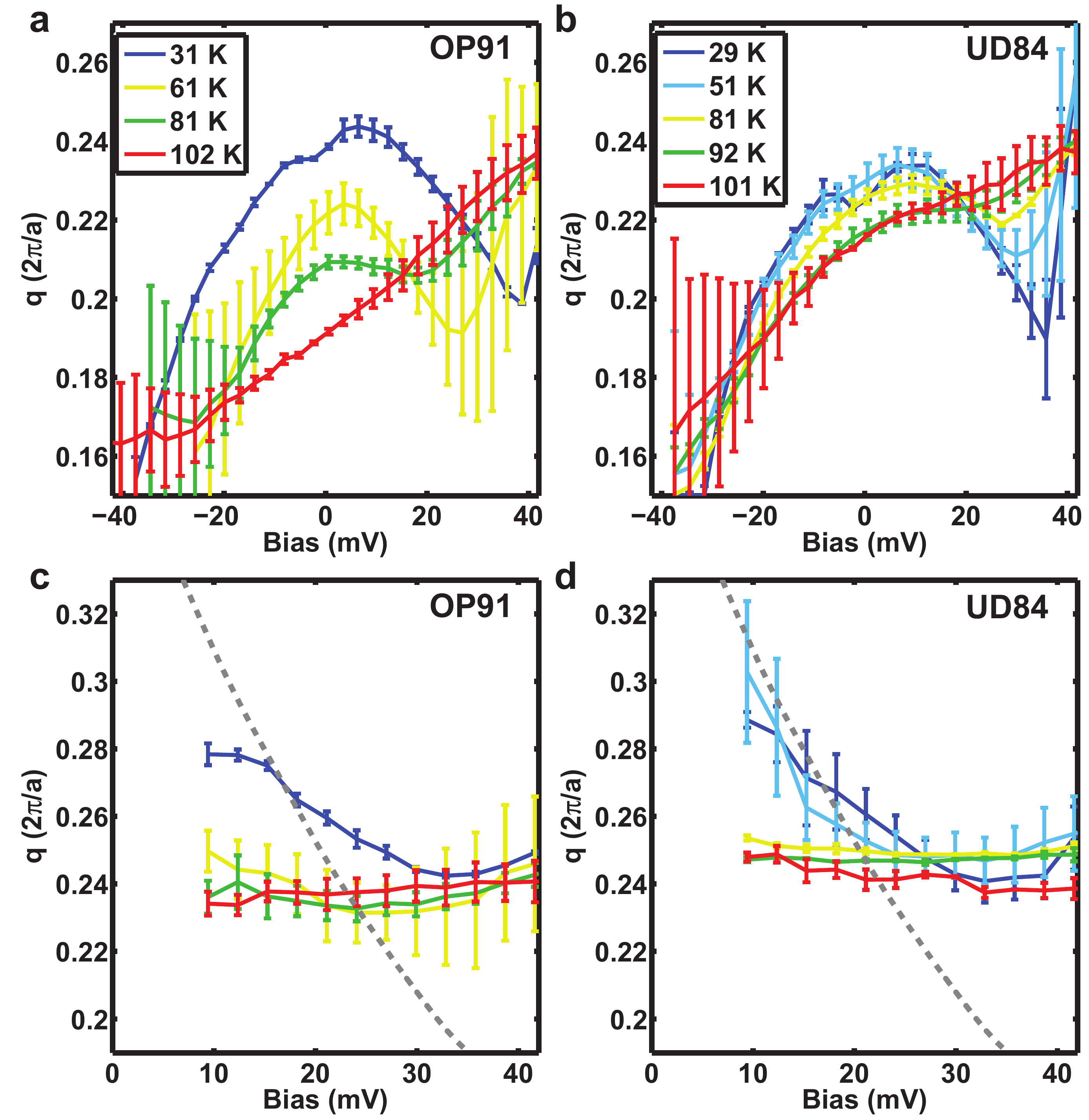}
\caption{\label{fig:2} $Q^*(E)$ for OP91 samples (a) and UD84 samples (b) at selected temperatures, measured along the Cu-O direction in units of $2\pi/a$, where $a$ is the nearest neighbor Cu-Cu distance. $Q^*(E)$ extracted from Z-map analysis for OP91 samples (c) and UD84 samples (d), for the same temperatures as in (a) and (b) respectively. Octet model prediction for $q_1$ is displayed as a gray dashed line in (c,d). The experimental data are reproduced from results reported in \cite{parker_fluctuating_2010}.}
\end{figure}

Figure \ref{fig:1}(a) shows an example of real space mapping of the low-energy conductance ($G(\bm{r},V)=dI/dV(\bm{r},V)$) using STM on an optimally doped (OP91, $T_c=91$ K) Bi-2212 sample, carried out at $35$ K, below $T_c$. Discrete Fourier transforms (DFT) of such conductance maps show strong peaks at wavevectors marked $Q^*$ along the Cu-O bond direction (Fig. \ref{fig:1}(b)). As we have previously reported, the $Q^*$ peaks are present in the DFT maps above $T_c$ and up to the pseudogap temperature $T^*$ \cite{parker_fluctuating_2010}. Despite the lack of sensitivity of the $Q^*$ intensity to the onset of superconductivity \cite{parker_fluctuating_2010}, we find these modulation to show systematic changes in their energy dependence with temperature. As shown in Fig. \ref{fig:2}(a,b), for both optimally doped (OP91) and underdoped (UD84, $T_c=84$ K) samples, the $Q^*$ wavevector disperses with energy, showing a particle-hole symmetric dispersion at temperatures well below $T_c$, with the $Q^*$ wavevector being approximately symmetric between positive and negative sample bias. At higher temperatures, specifically for $T > T_c$, we see that the energy dispersion, $Q^*(E)$, is no longer particle-hole symmetric. While the $Q^*$ modulation is present through the entire temperature range below the pseudogap temperature $T^*$, the particle-hole symmetry (asymmetry) of its dispersion correlates with the presence (absence) of superconductivity below (above) $T_c$.

In the superconducting state, well below $T_c$, we find that ratios of conductance maps $Z(\bm{r},V) = G(\bm{r},+V)/G(\bm{r},-V)$ \cite{hanaguri_quasiparticle_2007} provide an effective way of enhancing features of the data associated with superconductivity. As previously demonstrated these features of Z-map (marked as $q$\textquoteright s, Fig. \ref{fig:1}(c)) are similar to those expected from the so-called octet model for scattering of Bogoliubov-de Gennes quasiparticles (BdG-QPI) from random impurities in a d-wave superconductor (schematically shown in inset of Fig. \ref{fig:1}(c)) \cite{hoffman_imaging_2002, kohsaka_how_2008, hanaguri_quasiparticle_2007, wang_quasiparticle_2003, mcelroy_relating_2003}.
For temperatures well below $T_c$, $Q^*(E)$ from the Z-maps exhibits energy-dispersion trends that are in poor quantitative agreement with the the $q_1$ vector of the octet model extracted from the band structure obtained from angle-resolved photoemission (ARPES) measurements \cite{kivelson_how_2003, norman_phenomenological_1995} (dashed gray line in Fig. \ref{fig:2}(c,d)). However, by increasing the temperature above $T_c$, where $Q^*(E)$ is no longer particle-hole symmetric in the conductance maps, we find an artificial flattening of $Q^*(E)$ in the Z-map analysis (Fig. \ref{fig:2}(c,d)). This is evident since use of the Z-map assumes that the tunneling excitations measured by the STM tip are particle-hole symmetric BdG quasiparticles \cite{fujita_bogoliubov_2008}, a condition true only in the superconducting state (Fig. \ref{fig:2}(a,b)). Fluctuating superconductivity also cannot explain the behavior of $Q^*(E)$ above $T_c$, since such scenario would as well require the presence of particle-hole symmetry in $Q^*(E)$ — in clear disagreement with our observations (Fig. \ref{fig:2}). Therefore, to determine the nature of the $Q^*$ modulations not only below $T_c$ (particle-hole symmetric $Q^*(E)$) but also above (particle-hole asymmetric $Q^*(E)$), we focus only on information obtained from the conductance maps and their temperature dependence for the rest of this paper.

To obtain a unified understanding of the $Q^*$ modulations as a function of temperature, we calculate the quasiparticle scattering from a single impurity with a T-matrix approach \cite{wang_quasiparticle_2003}, in both the superconducting and the normal states, and consider how the results of such scattering would be modified in the presence of short range stripe order \cite{kivelson_how_2003, polkovnikov_spin_2003}. In this approach, the LDOS modulations due to impurity scattering are initially determined by 
\begin{eqnarray}
\chi_{0} ( \bm{q}, \omega) = \frac{1}{2\pi} Im[A_{11}(\bm{q}, \omega)+A_{22}(\bm{q}, -\omega)]
\end{eqnarray}
where $A(\bm{q}, \omega)$ is a $2\times2$ matrix, as prescribed by the Nambu-Gor\textquoteright kov spinor formalism:
\begin{eqnarray}
A(\bm{q}, \omega) = \int \frac{d^2 \bm{k}}{(2\pi)^2} G_0(\bm{k}+\bm{q}, \omega)T(\omega)G_0(\bm{k}, \omega)
\end{eqnarray}
For a single impurity, the T-matrix is
\begin{eqnarray}
T^{-1}(\omega) = (V_s \sigma_{3} + V_m I)^{-1} - \int \frac{d^2 \bm{k}}{(2\pi)^2}G_0(\bm{k},\omega)
\end{eqnarray}
where $G_0$ is the unperturbed ($2\times2$) Nambu-Gor\textquoteright kov single particle Green\textquoteright s function and $V_m$ ($V_s$) is the spin-flip (non-spin-flip) component of the impurity potential. 

\begin{figure}
\includegraphics[width=82mm]{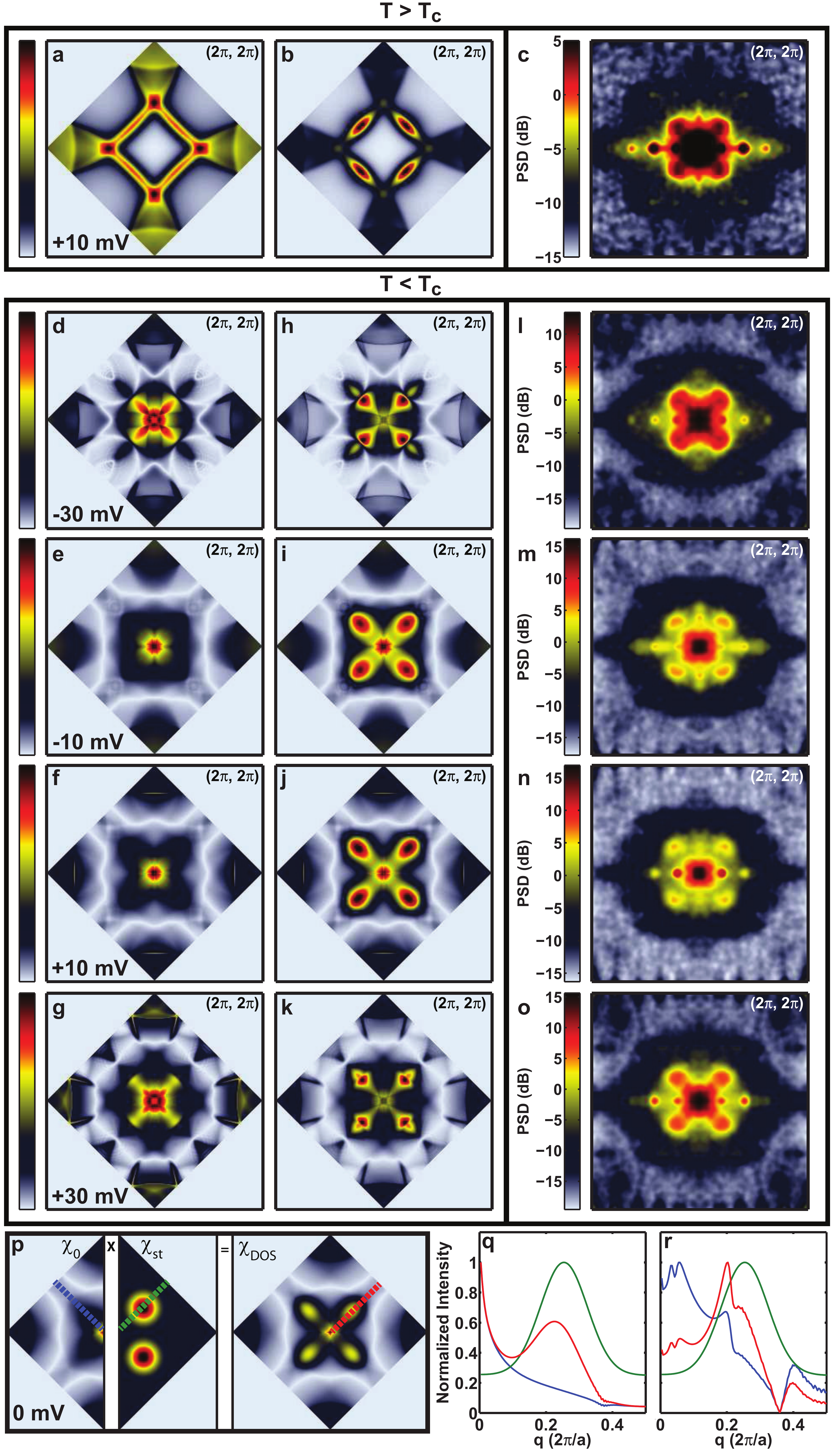}
\caption{\label{fig:3} (a) $\chi_{0}(\bm{q}, \omega = 10$ mV$)$ calculated for $\Delta_{\bm{k}} = 0$. (b) $\chi_{DOS}(\bm{q}, \omega)$ calculated from Eq. \ref{eq:5} using $\chi_{0}(\bm{q}, \omega)$ in (a). (c) DFT of $G(\bm{r}, V = 10$ mV$)$ on an OP91 sample at $T = 35$ K (setup bias and current equal $-150$ mV and $20$ pA, respectively). (d-g) $\chi_{0}(\bm{q}, \omega )$ calculated for $\Delta_{\bm{k}} = 45$ meV at the antinode, for selected energies. (h-k) $\chi_{DOS}(\bm{q}, \omega)$ calculated from $\chi_{0}(\bm{q}, \omega)$ in (d-g) respectively. (l-o) DFTs of $G(\bm{r}, V )$ on an OP91 sample at $T = 35$ K,  for selected energies (setup bias and current equal $-150$ mV and $40$ pA, respectively). (p) Schematic representation of Eq. \ref{eq:5}, where $\chi_{0}(\bm{q},\omega = 0$ mV$)$ is calculated in the superconducting case. (q) Line cuts through $\chi_0(\bm{q}, \omega)$ (blue), $\chi_{st}(\bm{q})$ (green), and $\chi_{DOS}(\bm{q}, \omega)$ (red) along the direction of $Q^*$ (see lines in (p)). To obtain figures (b, h-k, p) $\chi_{st}(\bm{q})$ was calculated from Eq. \ref{eq:6} for $\xi=0.1(2\pi/a)$, and $R=3$. (r) Similar line cuts for $\omega = 30$ mV ($\chi_0$ displayed in (g) and $\chi_{DOS}$ in (k)).}
\end{figure}

We assume the simplest form for the Green\textquoteright s function,
\begin{eqnarray}
G_0^{-1}(\bm{k},\omega) = (\omega + \imath \Gamma) I - \epsilon_{\bm{k}} \sigma_{3} - \Delta_{\bm{k}} \sigma_{1}
\end{eqnarray}
where $\epsilon_{\bm{k}}$ is the band structure obtained from photoemission experiments \cite{norman_phenomenological_1995}, $\Gamma$ (3 mV) represents quasiparticle broadening, $I$ is the identity matrix, $\sigma_1$ and $\sigma_3$ are Pauli matrices, and $\Delta_{\bm{k}}$ is the d-wave superconducting gap function. Above $T_c$, where  $\Delta_{\bm{k}}=0$, we find that the results of the above calculations, which are insensitive to the choice of the impurity potential ($V_m$, $V_s$), do not produce any peaks in the vicinity of $Q^*$ wavevectors within the experimental energy range (-40 mV to +40 mV) (see for example Fig. \ref{fig:3}(a)). Instead, the square-like contours seen in Fig. \ref{fig:3}(a) persist for all energies, although with varying intensities and wavelengths, in clear disagreement with the data (see for example Fig. \ref{fig:3}(c)). This behavior indicates that the modulations near $Q^*$ are not solely due to impurity scattering, rather as we have previously argued, a consequence of short-range stripe order in Bi-2212. 

To model the scattering from incipient short-range stripe order, we follow an approach proposed by Kivelson et al. \cite{kivelson_how_2003}. The presence of a short-range stripe order centered at $Q_{st}$, corresponding to a periodicity of $4a$, strongly enhances the quasiparticle scattering with momentum transfer near $Q_{st}$. While impurity scattering alone ($\chi_{0}$) might not produce peaks near $Q_{st}$, scattering from stripes will enhance any weak momentum- and energy-dependent features of $\chi_0$ near $Q_{st}$. Therefore, the LDOS measured by STM ($\chi_{DOS}(\bm{q}, \omega)$) will exhibit an energy dispersing peak, as long as $Q_{st}$ has finite correlation length (short-range).

The simplest phenomelogical way to model $\chi_{DOS}$ in the presence of scattering from this incipient stripe order is to write it as
\begin{eqnarray}
\label{eq:5}
\chi_{DOS}(\bm{q}, \omega) = \chi_{0}(\bm{q}) \times (\chi_{st}(\bm{q}, \omega) + 1)
\end{eqnarray}
where the stripe induced density of states takes the form
\begin{eqnarray}
\label{eq:6}
\chi_{st}(\bm{q}) = R\times exp\Bigg[{-\frac{(\bm{q}-\bm{Q}_{st})^2}{\xi^2}}\Bigg]
\end{eqnarray}
where $Q_{st} = (\pm 0.25, \pm 0.25) 2\pi/a_0$, $\xi$ determines the inverse correlation length and $R$ is the scattering enhancement at $Q_{st}$. This choice of ordering wavevector is also supported by the doping dependence of the $Q^*$ intensity we have previously reported \cite{parker_fluctuating_2010}. The result of this multiplication is a stripe-induced enhancement of the quasiparticle scattering near  $Q_{st}$  (e.g., Fig. \ref{fig:3}(b)) in good agreement with the data (Fig. \ref{fig:3}(c)). Note that the strong peak in the center of the data is due to long-wavelength inhomegeneities, which are absent in the model calculation.

The superconducting case below $T_c$, with $\Delta_{\bm{k}} = \Delta_0 [\cos(\bm{k}_x)-\cos(\bm{k}_y)]/2$, yields similar results, where in the absence of stripe order, $\chi_{0}$ does not produce peaks near $Q^*$ for low energies ($-20$ mV$<E<10$ mV, see for example Figs. \ref{fig:3}(e,f)) in disagreement with the data (Figs. \ref{fig:3}(l-o)). However, in the presence of stripe order $\chi_{DOS}$ reproduces the peaks near $Q^*$ for all energies (e.g., Figs. \ref{fig:3}(h-k)), in good agreement with the experimental data (Figs. \ref{fig:3}(l-o)). Note that the momentum structure of $\chi_{0}$ is encompassed by $\chi_{DOS}$ due to the short-range (wide) nature of $Q_{st}$ (Figs. \ref{fig:3}(q,r)).

\begin{figure}
\includegraphics[width=82mm]{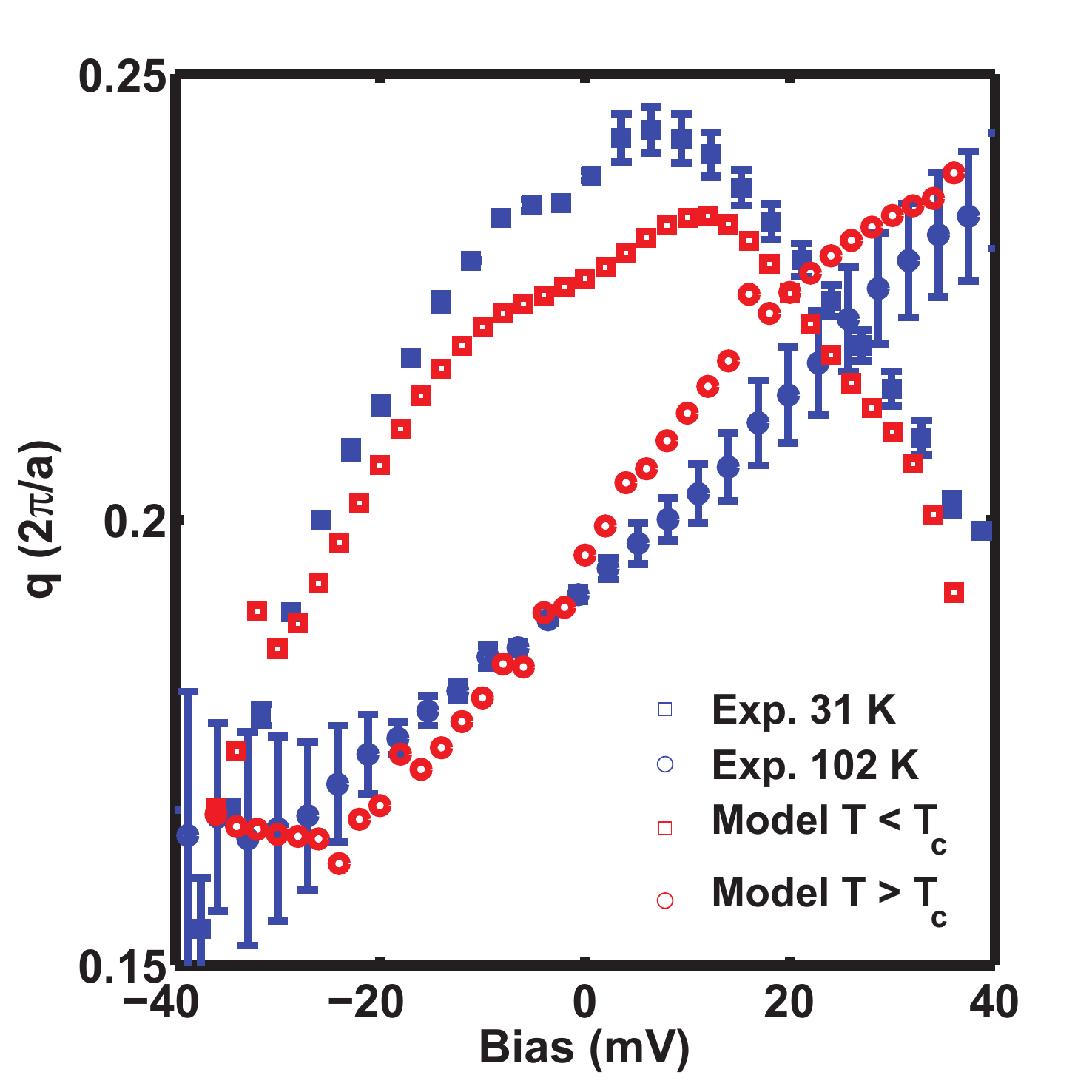}
\caption{\label{fig:4} Comparison of $Q^*(E)$, above and below $T_c$, obtained by the same fitting procedure for the model calculation and the experimental data measured for OP91. $Q^*(E)$ is measured in units of $2\pi/a$ where $a$ is the nearest neighbor Cu-Cu distance. }
\end{figure}

To quantitatively extract the energy dispersion of $Q^*(E)$ from $\chi_{DOS}$, we use the same fitting procedure used on the experimental data. Remarkably, $Q^*(E)$ extracted from our model calculation disperses both above and below $T_c$ and shows an excellent quantitative agreement with the experimental data for near optimal doping (Fig. \ref{fig:4}). Inclusion of both the dispersive feature of the hole-like band structure as well as an incipient stripe order appears to be required to capture the behavior of $Q^*(E)$.

In contrast to the data and calculation above $T_c$, the presence of a superconducting d-wave gap results in particle-hole symmetry near zero bias below $T_c$. Although the choice of scattering potential parameter does not make any difference in our results for $T > T_c$, we find that below $T_c$ the best agreement with the experimental data is obtained for $V_m=100$ mV, $V_s=0$ mV. This choice of parameters is similar to the one initially used to justify the use of octet model below $T_c$ \cite{wang_quasiparticle_2003} and is also supported by experimental measurements of scattering in a magnetic field \cite{hanaguri_coherence_2009}.

The success in quantitative modeling of the measurements of the energy dependence of $Q^*(E)$ modulations reported here demonstrates the importance of incipient short-range stripe order in understanding the electronic structure of the pseudogap phase and extends its relevance to the entire cuprate family, beyond the La-based compounds.  In this model the energy dependence of $Q^*(E)$ is determined by the symmetries of the electronic band structure. It therefore clarifies not only the presence of these modulations in the superconducting and normal states but also their dramatic distinction across $T_c$. The emerging physical picture is that although excitations of the high temperature cuprate Bi-2212 system have quasiparticle-like characteristics rising from the ARPES measured band structures below and above $T_c$, they also strongly scatter from a stripe-like incipient order that appears to develop below the pseudogap temperature $T^*$.

\bibliographystyle{h-physrev}
\bibliography{EHDA-lib}
\end{document}